\documentclass[10pt,twocolumn,letterpaper]{article}

\usepackage{cvpr}              
\usepackage{multirow}

\usepackage[dvipsnames]{xcolor}

\definecolor{cvprblue}{rgb}{0.21,0.49,0.74}
\usepackage[pagebackref,breaklinks,colorlinks,citecolor=cvprblue]{hyperref}

\def\paperID{10148} 
\def\confName{CVPR}
\def\confYear{2024}

\title{Purified and Unified Steganographic Network}

\author{Guobiao Li \thanks{Equal contribution} , Sheng Li \footnotemark[1] , Zicong Luo, Zhenxing Qian \thanks{Corresponding authors} , Xinpeng Zhang\footnotemark[2]\\
School of Computer Science, Fudan University\\
\{gbli20, lisheng, zcluo21, zxqian, zhangxinpeng\}@fudan.edu.cn
}

\begin{document}
\maketitle
\begin{abstract}
Steganography is the art of hiding secret data into the cover media for covert communication. In recent years, more and more deep neural network (DNN)-based steganographic schemes are proposed to train steganographic networks for secret embedding and recovery, which are shown to be promising. Compared with the handcrafted steganographic tools, steganographic networks tend to be large in size. It raises concerns on how to imperceptibly and effectively transmit these networks to the sender and receiver to facilitate the covert communication. To address this issue, we propose in this paper a Purified and Unified Steganographic Network (PUSNet). It performs an ordinary machine learning task in a purified network, which could be triggered into steganographic networks for secret embedding or recovery using different keys. We formulate the construction of the PUSNet into a sparse weight filling problem to flexibly switch between the purified and steganographic networks. We further instantiate our PUSNet as an image denoising network with two steganographic networks concealed for secret image embedding and recovery. Comprehensive experiments demonstrate that our PUSNet achieves good performance on secret image embedding, secret image recovery, and image denoising in a single architecture. It is also shown to be capable of imperceptibly carrying the steganographic networks in a purified network. 
Code is available at \url{https://github.com/albblgb/PUSNet} 
\end{abstract}    

\vspace{-2mm}
\section{Introduction}
Steganography aims to conceal secret data into a cover media, \textit{e.g.}, image\cite{kadhim2019comprehensive}, video \cite{mstafa2017compressed} or text \cite{majeed2021review}, which is one of the main techniques for covert communication through public channels. To conceal the presence of the covert communication, the stego media  (i.e., the media with hidden data) is required to be indistinguishable from the cover media. Early steganographic approaches \cite{filler2011minimizing, yao2015defining, zhang2016steganalytic} 
\begin{figure}[ht]
  \centering
  \includegraphics[width=0.95\linewidth]{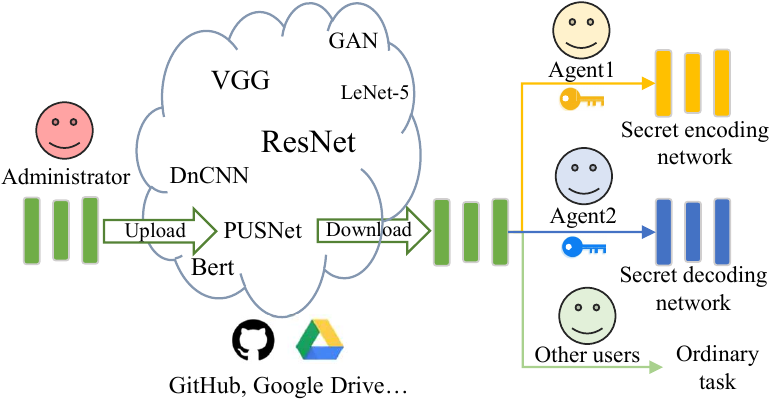} 
  \vspace{-0.1cm}
  \caption{Administrator covertly transmits the secret steganographic networks to agents using the proposed PUSNet.}
  \label{fig:nt}
\end{figure}
\begin{figure}[ht]
  \centering
  \includegraphics[width=0.92\linewidth]{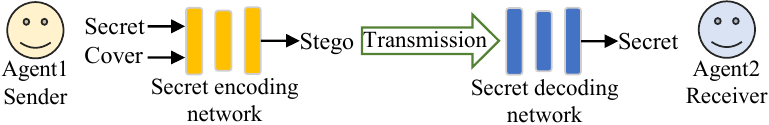} 
  \vspace{-0.1cm}
  \caption{The agents (i.e., sender and receiver) perform the covert communication task using the received steganographic networks.}
  \vspace{-3mm}
  \label{fig:st}
\end{figure}
are conducted under a handcrafted and adaptive coding strategy  
to minimize the distortion caused by data embedding.

In recent years, more and more deep neural network (DNN)-based steganographic schemes \cite{zhu2018hidden, baluja2019hiding, zhang2020udh, jing2021hinet, xu2022robust} are proposed to improve the steganographic performance.

A DNN-based steganographic scheme usually contains two main components, including a secret encoding (embedding) network and a secret decoding (recovery) network. The encoding network takes the cover media and the secret data as inputs to generate the stego media, while the decoding network retrieves the secrets from the stego media. These two networks are jointly learnt for optimized steganographic performance, which are shown to be superior to handcrafted steganographic tools.

Regardless of the steganographic schemes, we have to transmit the steganographic tools to the sender and receiver for secret embedding and recovery. This is not a trivial problem, especially for the DNN-based steganographic schemes which significantly increase the size of steganographic tools. A typical encoding or decoding network would occupy over 100MB of storage, which is much larger than handcrafted steganographic tools. It raises concerns on how we could covertly and effectively transmit the DNN-based steganographic tools to the sender and receiver for covert communication.



A promising solution to address the aforementioned issue is DNN model steganography, which has the capability to embed a secret DNN model into a benign DNN model without being noticed. The research of DNN model steganography is still in its infancy. Salem \textit{et al.} \cite{salem2022get} propose to establish a single DNN for both ordinary and secret tasks using multi-task learning. This scheme is not able to prevent unauthorized recovery of the secret DNN model from the stego DNN model (i.e., the model with hidden secret DNN model). Anyone who can access the stego DNN model would be able to perform both the ordinary and secret tasks. To deal with this issue, Li \textit{et al.} \cite{Li_Li_Li_Zhang_Qian_2023} propose to embed a steganographic network into a benign DNN model according to some side information to form a stego DNN model. The steganographic network can be restored from the stego DNN model only for authorized people who own the side information. Unfortunately, this scheme is tailored for concealing a secret decoding network. It remains unanswered regarding how we could imperceptibly and securely embed a secret encoding network into a benign DNN model. On the other hand, it requires the transmission of side information to the receiver for the recovery of the secret decoding network, which is not convenient in real-world applications.



In this paper, we try to tackle the problem of DNN model steganography by a Purified and Unified Steganographic Network (PUSNet). As shown in Fig. \ref{fig:nt}, our PUSNet is a purified network (i.e., the benign DNN model) that performs an ordinary machine learning task, which could be uploaded to the public DNN model repository by administrator. Agents (i.e., the sender or receiver) can download the PUSNet, and trigger it into a secret encoding network or a decoding network using keys possessed by them, where the keys are different for triggering different networks. Other users (those without the key) could also download the PUSNet for an ordinary machine learning task. Subsequently, the sender and receiver engage in covert communication tasks using the restored secret steganographic networks, as depicted in Fig \ref{fig:st}. By using our PUSNet, we imperceptibly conceal the secret encoding and decoding networks into a purified network. There is no need to look for secure and complicated ways to share the steganographic networks between the administrator and the agents.

 %

To flexibly switch the function of the PUSNet between an ordinary machine learning task and the secret embedding or recovery task, we formulate the problem of constructing the PUSNet in a sparse weight filling manner. In particular, we consider the purified network as a sparse network and the steganographic networks as the corresponding dense versions. We use a key to generate a set of weights to fill the sparse weights in the purified network to trigger a secret encoding or decoding network. As an instantiation, we design and adopt a sparse image denoising network as the purified network for concealing two steganographic networks, including a secret image encoding network and a secret image decoding network. Various experiments demonstrate the advantage of our PUSNet for steganographic tasks. The main contributions are summarized below.

\begin{itemize}

    \item [1)] We propose a PUSNet that is able to conceal both the secret encoding and decoding networks into a single purified network.


    \item[2)] We design a novel key-based sparse weight filling strategy to construct the PUSNet, which is effective in preventing unauthorized recovery of the steganographic networks without the use of side information.
    
    \item [3)] We instantiate our PUSNet as a sparse image denoising network with two steganographic networks concealed for secret image embedding and recovery, which justifies the ability of our PUSNet to covertly transmit the steganographic networks. 
\end{itemize}

\section{Related works} 

\textbf{DNN-based Steganography.}
Most of the existing DNN-based Steganographic schemes are proposed by taking advantage of the encoder-decoder structure for data embedding and extraction. Hayes \textit{et al.} \cite{hayes2017generating} pioneer the research of such a technique, where the secrets are embedded into a cover image or extracted from a stego-image using an end-to-end learnable DNN (i.e., a secret encoding or decoding network). Zhu \textit{et al.} \cite{zhu2018hidden} insert adaptive noise layers between the secret encoding and decoding network to improve the robustness. 
Baluja \textit{et al.}\cite{baluja2017hiding, baluja2019hiding} propose to embed a secret color image into another one for large capacity data embedding, where an extra network is designed to convert the secret image into feature maps before data embedding.
Zhang \textit{et al.} \cite{zhang2020udh} propose a universal network to transform the secret image into imperceptible high-frequency components, which could be directly combined with any cover image to form a stego-image.
Researchers also devote efforts to the design of invertible steganographic networks \cite{lu2021large, jing2021hinet, xu2022robust, guan2022deepmih}. 
Jing \textit{et al.} propose HiNet \cite{jing2021hinet, guan2022deepmih} to conceal the secrets into the discrete wavelet transform domain of a cover image using invertible neural networks (INN). Lu \textit{et al.} \cite{lu2021large} increase the channels in the secret branch of the INN to improve the capacity. Xu \textit{et al.} \cite{xu2022robust} introduce a conditional normalized flow to maintain the distribution of the high-frequency component of the secret image. 

\begin{figure*}[t]
\centering
  \includegraphics[width=0.84\linewidth]{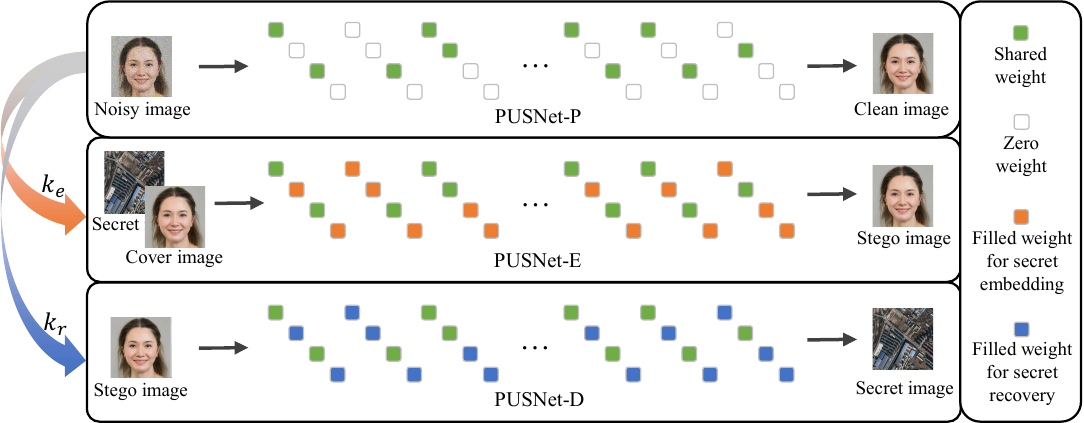}
  \vspace{-0.1cm}
  \caption{An overview of our proposed method.}
  \vspace{-0.3cm}
\label{fig:SWF}
\end{figure*}

\textbf{DNN Model Steganography.}
DNN model steganography aims to conceal a secret DNN model into a benign DNN model imperceptibly. The secret DNN models perform secret machine learning tasks, which are required to be covertly transmitted. While the benign/stego DNN models are released to the public for ordinary machine learning tasks. A few attempts have been made in literature for DNN model steganography. A straightforward strategy is to take advantage of the multi-task learning to train a single stego DNN model for both the ordinary and secret tasks \cite{salem2022get}. Such a strategy is not able to prevent unauthorized model extraction because anyone could use the stego DNN model for ordinary or secret tasks. Li \textit{et al.} \cite{Li_Li_Li_Zhang_Qian_2023} pioneer the work of embedding a steganographic network into a benign DNN model with the capability of preventing unauthorized model recovery. In this scheme, a partial of the neurons from the benign DNN model is carefully selected and replaced with those from the secret decoding network, while the rest neurons are learnt to construct a stego DNN model applicable to an ordinary machine learning task. The locations of the neurons of the secret decoding network in the stego DNN model are recorded as side information for model recovery. 
This scheme is tailored for embedding the secret decoding network, which is difficult to be adopted for the covert communication of secret encoding networks. Besides, the use of side information makes it inconvenient in real-world applications for model recovery.





\section{The proposed Method}
In this section, we elaborate in detail regarding how our PUSNet is established. We formulate the construction of the PUSNet as a sparse weight filling problem. Then, we introduce the loss function and training strategy for optimizing the PUSNet. Finally, we give the architecture of our PUSNet for instantiation.

\vspace{2mm}
\subsection{Sparse Weight Filling}
Our PUSNet is able to work on three different modes for an ordinary machine learning task, a secret embedding task, and a secret recovery task. In the following discussions, we denote our PUSNet as PUSNet-P, PUSNet-E, and PUSNet-D when it works as a purified network, secret encoding network, and secret decoding network, respectively.

Fig.~\ref{fig:SWF} gives an overview of how the PUSNet works on different modes. In particular, the PUSNet-P is a sparse network and the PUSNet-E and PUSNet-D are its dense versions. To switch the purified network to the steganographic networks, we have to fill the sparse weights in the PUSNet-P with new weights that are generated according to a key. 

Let's denote the PUSNet-P as $\mathbb{N}[\mathrm{W}\odot\mathrm{M}](\cdot) $, where $\mathbb{N}[\cdot]$ and $\mathrm{W}$ denote the architecture and weights of the network, respectively, $\odot$ represents the element-wise product and $\mathrm{M}$ is a binary mask with the same size as $\mathrm{W}$. We consider the image denoising task as an ordinary machine learning task for the PUSNet-P. Given a noisy image $\mathrm{x}_{no}$ and its clean version $\mathrm{x}_{cl}$, we can formulate the PUSNet-P by

\begin{equation}
    \mathbb{N}[\mathrm{W} \odot \mathrm{M}](\mathrm{x}_{no}) \to \mathrm{x}_{cl}.
\label{eq1}
\end{equation}

To switch the PUSNet-P into PUSNet-E, the sender could fill the sparse weights in the PUSNet-P by   
\begin{equation}
    \mathbb{N}[\mathrm{W} \odot \mathrm{M} + \mathrm{W}_{e} \odot \overline{\mathrm{M}}](\mathrm{x}_{co}, \mathrm{x}_{se}) \to \mathrm{x}_{st},
\label{eq2}
\end{equation}
where $\overline{\mathrm{M}}$ is a binary mask complementing $\mathrm{M}$, $\mathrm{x}_{co}$, $\mathrm{x}_{se}$ and $\mathrm{x}_{st}$ refer to the cover, secret and stego-image, $\mathrm{W}_e$ is a set of random weights initialized by 
\begin{equation}
    \mathrm{W}_e = \mathcal{I}(\mathbb{N}[\cdot], k_e),
\label{eq3}
\end{equation}
where $\mathcal{I}(\cdot)$ is an algorithm for seed (i.e., key) based weight initialization and $k_e$ is the key to trigger the secret encoding network. 
We use the Xavier \cite{glorot2010understanding} algorithm to initialize the filled weights in the implementation.

By the same token, the receiver could obtain the PUSNet-D by filling the sparse weights in the PUSNet-P by
\begin{equation}
    \mathbb{N}[\mathrm{W} \odot \mathrm{M} + \mathrm{W}_{r} \odot \overline{\mathrm{M}}](\mathrm{x}_{st}) \to \mathrm{x}_{se},
\label{eq4}
\end{equation}
where $\mathrm{W}_r$ a set of random weights initialized by 
\begin{equation}
\mathrm{W}_r = \mathcal{I}(\mathbb{N}[\cdot], k_r),
\end{equation}
where $k_r$ is a key to trigger the secret decoding network.



\begin{figure*}[t]
\centering
  \includegraphics[width=0.87\linewidth]{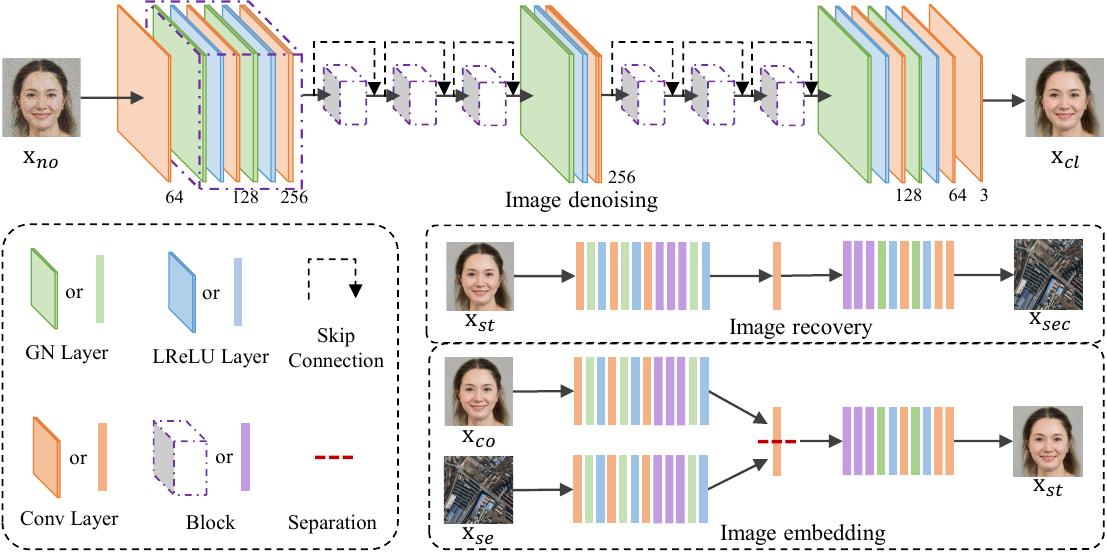}
  \caption{The architecture of the PUSNet.}
  \vspace{-0.3cm}
\label{fig:A}
\end{figure*}

\subsection{Loss Function}
To effectively train the PUSNet, we design the following loss terms, including an embedding loss, a recovery loss, and a purified loss. Next, we explain each loss term in detail.

\textbf{Embedding loss.}~
The embedding loss is designed to train the PUSNet-E which aims to embed a secret image $\mathrm{x}_{se}$ into a cover image $\mathrm{x}_{co}$ to generate a stego-image $\mathrm{x}_{st}$. It should be difficult for people to differentiate the stego-image from the cover image. To this end, we compute the embedding loss below:
\begin{equation}
     \mathcal{L}_{emb} = \sum^{N}_{n=1}\ell_2(\mathrm{x}_{st}^{n}, \mathrm{x}_{co}^{n}),
\label{eq6}
\end{equation}
where $\mathrm{x}_{co}^{n}$ and $\mathrm{x}_{st}^{n}$ are the $n$-th cover image and the corresponding stego-image in the training set, $N$ is the number of samples for training, and $\ell_2$ is the L-2 norm to measure the distortion between the cover and stego-image.

\textbf{Recovery loss.}~
The recovery loss is designed to train the PUSNet-D which recovers the secret image $\mathrm{x}_{se}$ from the stego-image $\mathrm{x}_{st}$. The recovered secret image should be close to $\mathrm{x}_{se}$. Therefore, the recovery loss is given by 

\begin{equation}
     \mathcal{L}_{rec} = \sum^{N}_{n=1}\ell_2(\mathrm{x}_{sec}^{n}, \mathrm{x}_{se}^{n}),
\label{eq7}
\end{equation}
where $\mathrm{x}_{st}^{n}$ and $\mathrm{x}_{sec}^{n}$ refer to the $n$-th secret image and the corresponding recovered version for training.


\textbf{Purified loss.}~ 
The purified loss is designed to train the PUSNet-P which conducts image denoising to restore a clean image from a noisy one. Given a noisy image $\mathrm{x}_{no}$ for input, the PUSNet-P is expected to output a restored image that is close to the clean version $\mathrm{x}_{cl}$. The denoising loss is given by
\begin{equation}
     \mathcal{L}_{den} = \sum^{N}_{n=1}\ell_2(\mathrm{x}_{d}^{n}, \mathrm{x}_{cl}^{n}),
\label{eq8}
\end{equation}
where $\mathrm{x}_{d}^{n}$ and $\mathrm{x}_{cl}^n$ refer to the $n$-th restored image and the corresponding ground-truth clean image for training.


During the training, we only update the sparse weights in the PUSNet-P (denoted as $\mathrm{W}_s$), which is shared among PUSNet-P, PUSNet-E and the PUSNet-D. Please refer to the green nodes in Fig. \ref{fig:SWF} for illustration of $\mathrm{W}_s$. In other words, we only update a partial of the weights in PUSNet (i.e., $\mathrm{W}_s$) to optimize the performance of the PUSNet-P, PUSNet-E, and PUSNet-D on different tasks. Let $\alpha$ be the learning rate, we update $\mathrm{W}_s$ below using gradient descent: 

\begin{equation}
     \mathrm{W}_s = \mathrm{W}_s - \alpha \ (\lambda_e \nabla_{\mathrm{W}_s} \mathcal{L}_{emb} + \lambda_r \nabla_{\mathrm{W}_s} \mathcal{L}_{rec} + \lambda_d \nabla_{\mathrm{W}_s} \mathcal{L}_{den} ),
\label{eq10}
\end{equation}
where $\lambda_e$, $\lambda_r$ and $\lambda_d$ are hyper-parameters for balancing the contributions of the gradients computed from PUSNet-E, PUSNet-D and PUSNet-P, respectively.


\subsection{Network Architecture}
Similar to classic denoising DNN models \cite{zhang2017beyond, mao2016image, zhang2018ffdnet}, our PUSNet is a DNN constructed by stacking convolutional (Conv), normalization, and activation layers. Fig.~\ref{fig:A} depicts the architecture of our PUSNet, which consists of 19 Conv layers. There is a group normalization (GN) \cite{wu2018group} layer and a Leaky Rectified Linear Unit (LReLU) \cite{he2015delving} before each Conv layer except for the first and last one. We adopt skip connections \cite{he2016deep} from the fourth to sixteenth Conv layers. By following the suggestion given in \cite{he2016identity}, we place the skip connections between the Conv and GN layers.

Taking a single image as input, the above architecture outputs a predicted image with the same size as the input, which is suitable for the image denoising and secret image recovery tasks. Since a secret image embedding task requires two images (i.e., $\mathrm{x}_{co}$ and $\mathrm{x}_{se}$) for input, we propose below an adaptive strategy to make the PUSNet suitable for secret encoding. 
The basic concept is to duplicate the first half of the network into two identical sub-networks to process $\mathrm{x}_{co}$ and $\mathrm{x}_{se}$ separately to obtain two feature maps from the cover and secret image. These two feature maps are then concatenated and fed into the second half of the network to generate the stego-image $\mathrm{x}_{st}$, as shown in the lower part in Fig.~\ref{fig:A}. In our implementation, we take layers before the tenth Conv layer from the PUSNet to extract two feature maps from $\mathrm{x}_{co}$ and $\mathrm{x}_{se}$. Then, we separate the filters of the tenth Conv layer into two halves, where each half is used to convolve with the features of $\mathrm{x}_{co}$ or $\mathrm{x}_{se}$. As such, we can directly concatenate the convolved features and feed them into the rest of the PUSNet to generate $\mathrm{x}_{st}$. Such a strategy enables the PUSNet to process multiple images without causing additional overhead, which improves the flexibility of the PUSNet for different tasks.

\begin{table*}[t]
\renewcommand{\arraystretch}{1.00} 
\caption{Performance comparisons on different datasets. ``$\uparrow$": the larger the better, ``$\downarrow$": the smaller the better.}
\vspace{-0.1cm}

\centering
\resizebox{1.0\linewidth}{!}{
\begin{tabular}{@{}c|cccc|cccc|cccc@{}}
    \hline
    \multirow{3}{*}{Methods}  & \multicolumn{12}{c}{Cover/Stego-image pair}  \\
    \cline{2-13}
    &\multicolumn{4}{c|}{DIV2K}   &  \multicolumn{4}{c|}{COCO} & \multicolumn{4}{c}{ImageNet}  \\ 
    \cline{2-13}
    &PSNR(dB)$\uparrow$&SSIM$\uparrow$&APD$\downarrow$&RMSE$\downarrow$ &PSNR(dB)$\uparrow$&SSIM$\uparrow$&APD$\downarrow$&RMSE$\downarrow$ &PSNR(dB)$\uparrow$&SSIM$\uparrow$&APD$\downarrow$&RMSE$\downarrow$ \\
    \cline{1-13}
    HiDDeN \cite{zhu2018hidden}& 28.19& 0.9287& 8.01& 11.00& 29.16& 0.9318& 6.91& 9.60& 28.87& 0.9234& 7.43& 10.21 \\
    Baluja \cite{baluja2019hiding}& 28.42& 0.9347& 7.92& 10.64& 29.32& 0.9374& 7.04& 9.36& 28.82& 0.9303& 7.68& 10.21 \\
    UDH \cite{zhang2020udh}& 37.58& 0.9629& 2.38& 3.40& 38.01& 0.9033& 6.12& 9.55& 37.89& 0.9559& 2.30& 3.29 \\
    HiNet \cite{jing2021hinet}& 44.86& 0.9922& 1.00& 1.53& 46.47& 0.9925& 0.81& 1.30& 46.88& 0.9920& 0.81& 1.26 \\
    PUSNet-E & 38.15& 0.9792& 2.30& 3.33& 39.09& 0.9772& 2.01& 2.96& 38.94& 0.9756& 2.21 & 3.06 \\

    \hline
    \multirow{3}{*}{Methods} & \multicolumn{12}{c}{Secret/Recovered image pair} \\
    \cline{2-13}
    &\multicolumn{4}{c|}{DIV2K}   &  \multicolumn{4}{c|}{COCO} & \multicolumn{4}{c}{ImageNet}  \\ 
    \cline{2-13}
    &PSNR(dB)$\uparrow$&SSIM$\uparrow$&APD$\downarrow$&RMSE$\downarrow$ &PSNR(dB)$\uparrow$&SSIM$\uparrow$&APD$\downarrow$&RMSE$\downarrow$ &PSNR(dB)$\uparrow$&SSIM$\uparrow$&APD$\downarrow$&RMSE$\downarrow$ \\
    \cline{1-13}
    HiDDeN \cite{zhu2018hidden}& 28.42& 0.8695& 7.62& 9.94& 28.81& 0.8576& 7.20& 9.54& 28.23& 0.8435& 7.83& 10.47 \\
    Baluja \cite{baluja2019hiding}& 28.53& 0.9036 & 7.53& 10.66& 29.13& 0.9091& 6.61& 9.80& 27.63& 0.8909& 8.33 & 12.26\\
    UDH \cite{zhang2020udh}& 30.52& 0.9120& 5.62& 7.92& 30.52& 0.9120& 5.62& 7.92& 29.63& 0.8916& 6.67& 10.33 \\
    HiNet \cite{jing2021hinet}& 28.66& 0.8507& 7.25& 9.68& 28.08& 0.8181& 7.80& 10.49& 27.94& 0.8159& 8.03& 10.83 \\
    PUSNet-D & 26.88& 0.8363& 8.75& 11.95& 26.96& 0.8211& 8.71& 12.14& 26.28& 0.8028& 9.58& 13.43 \\  
    \hline
\end{tabular}}
\vspace{-0.1cm}

\label{tab:comparison}
\end{table*}

\begin{figure*}[ht]
  \centering
  \includegraphics[width=0.96\linewidth]{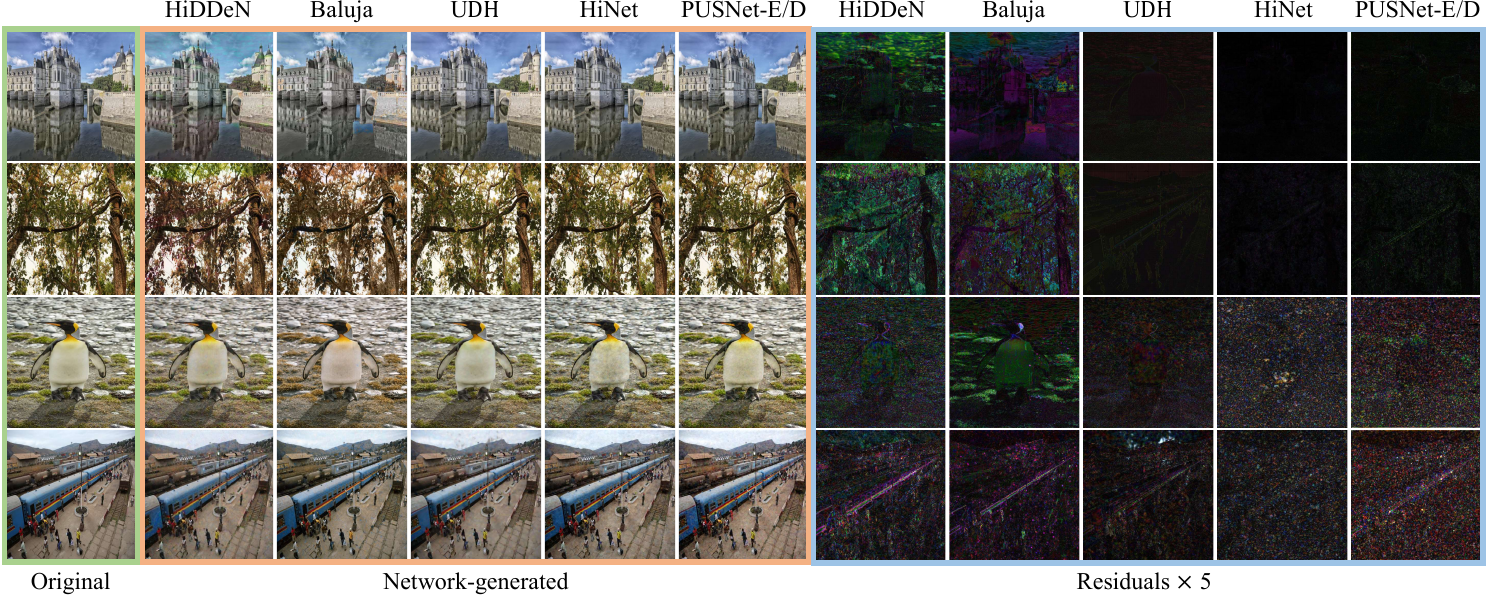}
  \caption{Examples of the stego and recovered images generated using different schemes, with a
green border on the original images, an orange border on the generated images, and a blue border on × 5 magnified residuals between them. The cover/stego and secret/recovered images are given in the first two rows and the last two rows, respectively.}
\vspace{-0.3cm}
  \label{fig:vsp}
\end{figure*}

\subsection{Sparse Mask Generation}
As what have mentioned before, the purified network (i.e., PUSNet-P) has to be a sparse network to trigger the secret encoder or decoder network (i.e., PUSNet-E and PUSNet-D). Next, we explain how we initialize a sparse network PUSNet-P based on the network architecture given in the previous section. Randomly generating a sparse mask (i.e., $\mathrm{M}$) may not be the best solution because it is weak in maintaining the performance of the purified network for image denoising.  

Fortunately, researchers have proposed several approaches to prune the networks at the initialization stage \cite{lee2018snip, wang2020picking, tanaka2020pruning, frankle2020pruning}, which are shown to be effective for initializing a sparse network. In our implementation, we adopt a magnitude-based pruning method to generate the sparse mask \cite{frankle2020pruning}. Given a sparse ratio $\mathcal{S}$ and the total number of weights in the PUSNet $\mathcal{N}$, we generate the sparse mask $\mathrm{M}$ as follows.

\begin{itemize}
   \item[1)] Initialize all the weights in the PUSNet using a random seed, and sort them in descending order.
    \item[2)] Compute $\mathrm{M}$ by
    \begin{equation}
     \mathrm{M} \gets 1_{(\mathrm{W}_{0} > t)},
     \label{eq11}
    \end{equation}
    where $1$ is the indicator function, $\mathrm{W}_{0}$ is the initialized weights, and $t$ is a threshold equals to the $p$-th largest weight in $\mathrm{W}_{0}$,where $p=\lfloor \mathcal{S}\cdot\mathcal{N} \rfloor$ and $\lfloor  \rfloor$ is the floor operation.
\end{itemize}

\section{Experiments}
\subsection{Implementation Details}

\textbf{Training.}~ Our PUSNet is trained on the DIV2K training dataset \cite{8014884}, which consists of 800 high-resolution images. We randomly crop 256 $\times$ 256 patches from the dataset and apply horizontal and vertical flipping for data augmentation. The mini-batch size is set to 8, with half of the patches randomly selected as the cover images and the remaining patches as the secret images. 
We add Gaussian noise into the patches to generate noisy images for training.
The PUSNet is trained for 3,000 iterations using Adam \cite{kingma2014adam} optimizer with default parameters using a fixed weight decay of $1 \times 10^{-5}$ and an initial learning rate of $1 \times 10^{-4}$. The learning rate is reduced by half every 500 iterations. The hype-parameters $\lambda_e$, $\lambda_r$ and $\lambda_{d}$ are set as 1.0, 0.75, 0.25, respectively. 
Unless stated otherwise, we set the sparse ratio as $\mathcal{S}=0.9$.

%

\textbf{Evaluation.}~We evaluate the performance of our PSUNet on three test sets, including the DIV2K test dataset, 1000 images randomly selected from the ImageNet test dataset \cite{russakovsky2015imagenet}, and 1000 images randomly selected from the COCO dataset \cite{lin2014microsoft}. All test images are resized to 512 × 512 pixels before being fed into the network. We adopt four metrics to measure the visual quality of the images, including Peak Signal-to-Noise Ratio (PSNR), Structural Similarity Index (SSIM) \cite{wang2004image}, Averaged Pixel-wise Discrepancy (APD), and Root Mean Square Error (RMSE). 
We adopt two steganalysis tools, including StegExpose \cite{DBLP:journals/corr/Boehm14} and SiaStegNet \cite{you2020siamese}, to evaluate the undetectability of the stego-images generated by PUSNet-E. We also employ three strategies to detect for DNN model steganalysis, which aim to detect the existence of secret DNN models (e.g., the steganographic networks) from a purified DNN model. The network-generated images are quantified before the evaluation. 
All our experiments are conducted on Ubuntu 18.04 with four NVIDIA RTX 3090 Ti GPUs.

\begin{table*}[t]
\setlength\tabcolsep{5pt}
\renewcommand{\arraystretch}{1.00} 
\centering
\caption{Comparison of the denoising performance of the PUSNet-P and PUSNet-C on different datasets.}
\vspace{-0.2cm}

\resizebox{1.\linewidth}{!}{
\begin{tabular}{c|cccc|cccc|cccc}
    \hline
    \multirow{2}{*}{Image pairs} &\multicolumn{4}{c|}{DIV2K}   &  \multicolumn{4}{c|}{COCO} & \multicolumn{4}{c}{ImageNet}  \\ 
    \cline{2-13}
    &PSNR(dB)$\uparrow$&SSIM$\uparrow$&APD$\downarrow$&RMSE$\downarrow$ &PSNR(dB)$\uparrow$&SSIM$\uparrow$&APD$\downarrow$&RMSE$\downarrow$ &PSNR(dB)$\uparrow$&SSIM$\uparrow$&APD$\downarrow$&RMSE$\downarrow$ \\
    \hline
    $\mathbf{x}_{no}$/$\mathbf{x}_{cl}$ & 22.11& 0.4432& 15.95& 19.99 & 22.11& 0.3907& 15.96& 20.00& 21.11& 0.3902& 15.96& 20.00 \\
    PUSNet-P$(\mathbf{x}_{no})$/$\mathbf{x}_{cl}$ & 32.25& 0.9080& 4.57& 6.37& 32.95& 0.8926& 4.29& 5.89& 32.96& 0.8922& 4.36& 5.94 \\
    PUSNet-C$(\mathbf{x}_{no})$/$\mathbf{x}_{cl}$ & 33.03& 0.9236& 4.11& 5.84& 33.68& 0.9074& 3.92& 5.54& 33.66& 0.9073& 4.00& 5.52 \\
    \hline
\end{tabular}}
\vspace{-0.2cm}

\label{tab:BP}
\end{table*}

\subsection{Visual quality}

We evaluate the visual quality of the stego-image and the recovered secret image (termed as the recovered image for short) using our PUSNet. For better assessment, we compare our PSUNet against several state-of-the-art (SOTA) DNN-based steganographic techniques, including Baluja \cite{baluja2019hiding}, HiDDeN \cite{zhu2018hidden}, UDH \cite{zhang2020udh}, and HiNet \cite{jing2021hinet}. For fairness, we retrain the aforementioned models on the DIV2K training dataset and evaluate their performance under the same settings as ours. As given in Table \ref{tab:comparison}, we can see that our PUSNet outperforms Baluja \cite{baluja2019hiding} and HiDDeN \cite{zhu2018hidden} in all four metrics in terms of the visual quality of the stego-images. Specifically, the PUSNet-E achieves over 9.73 dB, 9.77 dB, and 10.07 dB performance gain on DIV2K, COCO, and ImageNet, respectively. While the PUSNet-D does not perform as well as the PUSNet-E, with a PSNR of over 26dB for the recovered images, which is still acceptable for revealing sufficient content in the recovered images.



Fig.~\ref{fig:vsp} illustrates the stego and recovered images using different schemes. To highlight the difference between the cover/stego or secret/recovered image pairs, we magnify their residuals by 5 times. It can be seen that our PUSNet-E and PUSNet-D are able to generate stego and recovered images with high visual quality. By using PUSNet-E, the residual between the stego-image and the cover image is at a low visual level, which is the second best among all the schemes. By using the PUSNet-D, we observe noticeable noise in the residual between the secret and recovered images. But we could still be able to look into the details of the image content from the recovered image. Overall, our PUSNet is capable to be served as a steganographic tool for covert communication.  


\subsection{Undetectability of the Stego-images}
Next, we evaluate the undetectability of the stego-images generated using our PUSNet-E. We use two popular image steganalysis tools that are publicly available to carry out the evaluation, including StegExpose \cite{DBLP:journals/corr/Boehm14} and SiaStegNet \cite{you2020siamese}. The former is a traditional steganalysis tool which assembles a set of statistical methods, while the latter is a DNN-based steganalysis tool.

\begin{figure}
\vspace{-1mm}
  \begin{minipage}[t]{0.48\linewidth}
    \centering
    \includegraphics[width=0.96\textwidth,height=1.08\textwidth]{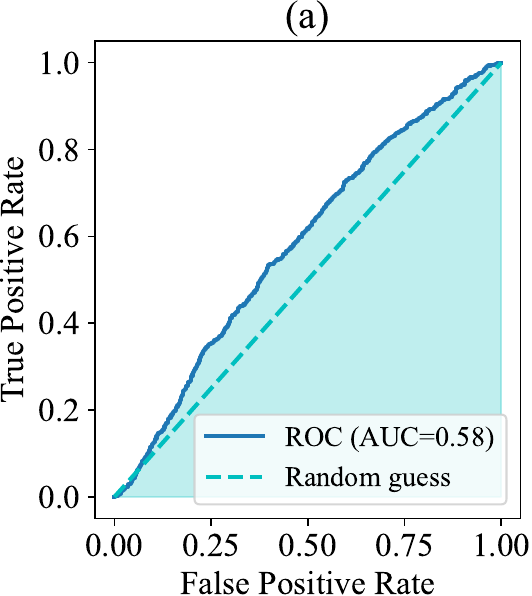}
    
  \end{minipage}%
  \begin{minipage}[t]{0.48\linewidth}
    \centering
    \includegraphics[width=0.96\textwidth,height=1.08\textwidth]{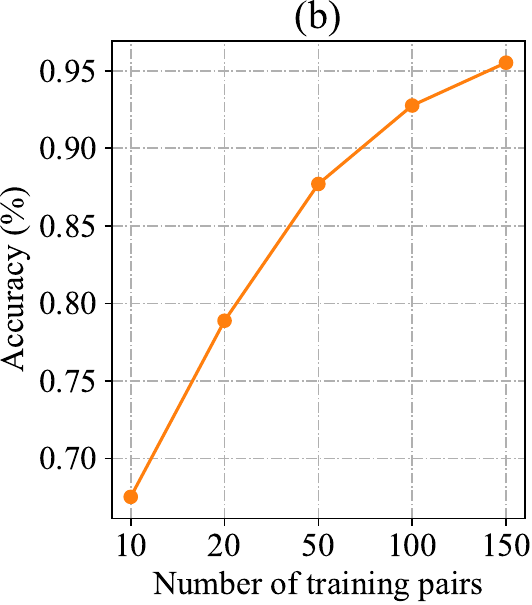}
  \end{minipage}
    \caption{The undetectability of the stego-images generated using PUSNet-E against (a) StegExpose and (b) SiaStegNet.}
    \label{fig:safi}
    \vspace{-0.2cm}
    
\end{figure}

We follow the same protocol as in \cite{jing2021hinet} to use the StegExpose. In particular, we use our PUSNet-E on all the cover images in the three testing datasets to generate the stego-images, which are then fed into the StegExpose for evaluation. We obtain a receiver operating characteristic (ROC) curve by varying the detection thresholds in StegExpose, which is shown in Fig.~\ref{fig:safi} (a). The value of area under curve (AUC) of this ROC curve is 0.58, which is very close to random guessing (AUC=0.5). This demonstrates the high undetectability of our stego-images against the StegExpose.

In order to conduct the evaluations using the SiaStegNet, we follow the protocol given in \cite{guan2022deepmih} to train SiaStegNet using different numbers of cover/stego-image pairs to investigate how many image pairs are needed to make SiaStegNet capable to detect the stego-images. Fig.~\ref{fig:safi}(b) plots the detection accuracy of the SiaStegNet by varying the number of image pairs for training. It can be observed that, in order to accurately detect the existence of secret data, the adversary needs to collect at least 100 labeled cover/stego-image pairs. This could be challenging in real-world applications. Since there is always a trade-off between the payload and undetectability \cite{zhu2021destroying, zhu2022image}, the sender could reduce the amount of the payload of the secret information in a stego-image to improve the undetectability. 

\begin{table*}[t]
\setlength\tabcolsep{5pt}
\renewcommand{\arraystretch}{1.0} 
\centering
\caption{Steganographic performance (mean$\pm$std) of the PUSNet-ER and PUSNet-DR.}
\vspace{-0.1cm}

\resizebox{0.90\linewidth}{!}{
\begin{tabular}{c|cc|cc|cc}
    \hline
    \multirow{2}{*}{Image pairs} &\multicolumn{2}{c|}{DIV2K}   &  \multicolumn{2}{c|}{COCO} & \multicolumn{2}{c}{ImageNet}  \\ 
    \cline{2-7}
    &PSNR(dB)$\uparrow$&APD$\downarrow$&PSNR(dB)$\uparrow$&APD$\downarrow$&PSNR(dB)$\uparrow$&APD$\downarrow$\\
    \hline
    PUSNet-ER$(\mathbf{x}_{se}, \mathbf{x}_{co})$/$\mathbf{x}_{co}$ & 8.81$\pm$1.74& 83.06$\pm$17.79 & 8.01$\pm$2.07& 92.19$\pm$22.33& 7.74$\pm$1.89& 95.45$\pm$20.54  \\
    PUSNet-DR$(\mathbf{x}_{st})$/$\mathbf{x}_{se}$ &6.52$\pm$0.78& 107.06$\pm$11.07& 6.25$\pm$0.97 &107.01$\pm$14.03& 6.73$\pm$0.83& 100.30$\pm$10.78 \\
    \hline
\end{tabular}}
\vspace{-0.1cm}

\label{tab:rsp}
\end{table*}

\subsection{Undetectability of the DNN model}

Since we try to imperceptibly conceal the steganographic networks into a purified network, it is necessary to conduct an analysis to detect the existence of secret DNN models in a purified model that is transmitted through public channels. We term such a task as the DNN model steganalysis. It is unfortunate that almost all the existing steganalysis tools are designed for media (image/video/text) steganalysis. 
In this section, we empirically adopt several strategies for DNN model steganalysis. We assume that the adversary possesses the PUSNet-C, which is trained only for image denoising using the DIV2K training dataset. 
\begin{figure}[ht]
  \centering
  \includegraphics[width=0.96\linewidth]{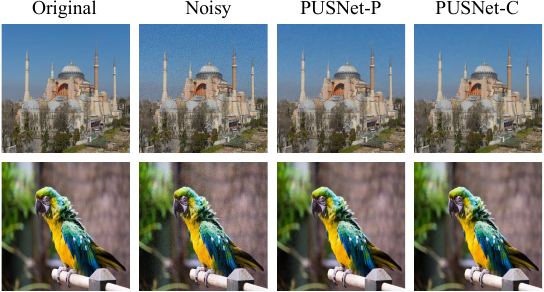}
  \caption{Visual comparisons of denoised images of the PUSNet-P and PUSNet-C. From left to right, the original clean images, the noisy images, the denoised images using PUSNet-P, and the denoised images using PUSNet-C.}
  \label{fig:DP}
\end{figure}
The PUSNet-C is regarded as the pure purified model, which does not contain any secret networks. Its counterpart is the PUSNet-P which can be used to trigger the PUSNet-E and PUSNet-D. We conduct the DNN model steganalysis in the following three aspects.

\textbf{Performance reduction.}~  
In this strategy, we aim to measure the performance reduction between the PUSNet-P and PUSNet-C on the image denoising task. The PUSNet-P should have similar denoising ability compared to the PUSNet-C to avoid being noticed.
Table.~\ref{tab:BP} provides the visual quality between the denoised/clean image pairs using different models, where PUSNet-P$(\mathbf{x}_{no})$ and PUSNet-C$(\mathbf{x}_{no})$ represent denoised versions of the image $\mathbf{x}_{no}$ using the PUSNet-P and PUSNet-C, respectively. It can be seen that both the PUSNet-P and PUSNet-C are equipped with good image denoising ability, which significantly improve the visual quality of the images after denoising. Compared with the PUSNet-C, the performance reduction of PUSNet-P is neglectable, with less than 0.8dB decrease in PSNR on the COCO dataset. Figure~\ref{fig:DP} visualizes some examples of denoised images using the PUSNet-P and PUSNet-C, where we can hardly observe the difference between the denoised images using different denoising models.

\begin{figure}[t]
\centering
\vspace{-0.1cm}
  \includegraphics[width=0.58\linewidth]{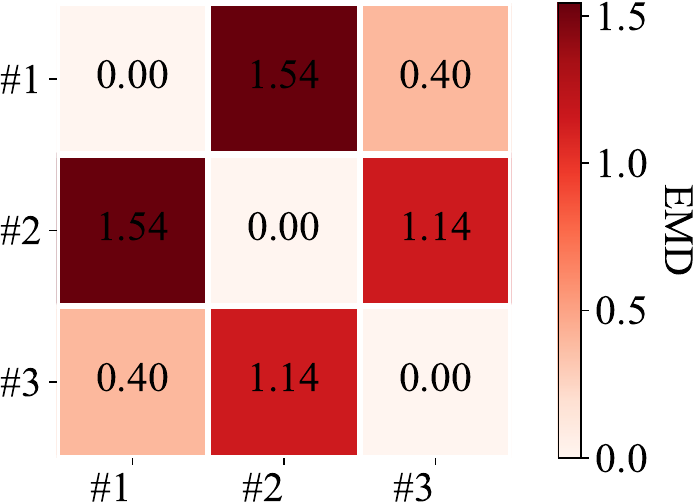}
  \caption{The pairwise EMDs among the PUSNet-C$_1$ (\#1), PUSNet-C$_2$ (\#2) and PUSNet-P (\#3).}
\vspace{-0.2cm}
\label{fig:WD}
\end{figure}

\textbf{Weight Distribution.}
In this strategy, we aim to measure the distance between the distributions of the weights in PUSNet-P and PUSNet-C. We believe such a distance could be useful for DNN model steganalysis.    
We adopt the Earth Mover’s Distance (EMD) \cite{rubner2000earth}
to measure the distance between weight distributions of the PUSNet-P and PUSNet-C. 
Here, we provide two versions of PUSNet-C, including PUSNet-C$_1$ and PUSNet-C$_2$, which are trained using slightly different strategies. Specifically, the weight decays of their optimizers are set as $1 \times 10^{-5}$ and 0 respectively.
We consider the PUSNet-P to be secure if the EMDs between the PUSNet-P and PUSNet-C$_1$ / PUSNet-C$_2$ is less than that between PUSNet-C$_1$ and PUSNet-C$_2$.
Fig.~\ref{fig:WD} plots the pairwise EMDs among different model pairs using POT \cite{flamary2021pot}. It can be seen that the weight distributions of the PUSNet-P and PUSNet-C$_1$ are similar as evidenced by a low EMD. Moreover, the EMD between PUSNet-P and PUSNet-C$_2$ does not exceed the EMD between PUSNet-C$_1$ and PUSNet-C$_2$. Therefore, it is difficult to determine the existence of the secret modals in our PUSNet-P by computing the distance of the weight distribution between PUSNet-P and PUSNet-C. 

\textbf{Analysis of steganographic networks Leakage.}~
One may also wonder about the possibility of secret steganographic networks leakage if an adversary launches the sparse weight filling on the PUSNet-P by using a key that is randomly guessed. In what follows, we evaluate if it is possible to leak the PUSNet-E and PUSNet-D from the PUSNet-P under such an attack.
We conduct the above sparse weight filling attack 1000 times to see if the PUSNet-E and PUSNet-D can be successfully triggered from the PUSNet-P. Table.~\ref{tab:rsp} reports the PSNR and APD of the stego and recovered images generated using the randomly triggered networks, where PUSNet-ER and PUSNet-DR refer to the randomly triggered secret encoder and decoding networks, respectively. It can be seen that, both the stego and recovered images are poor in visual quality on different datasets, where the PSNR is less than 9dB and the APD is over 80. This indicates that it is difficult for the attacker to launch a successful attack by using a random key to trigger the secret encoding and decoding network.


    

\begin{table}[t]
\renewcommand{\arraystretch}{1.0} 
\centering
\vspace{-1mm}
\caption{Performance comparisons on hiding steganographic networks. $\searrow$: performance reduction on the task.}
\resizebox{1.\linewidth}{!}{
\begin{tabular}{@{}c|cc|cc@{}}
    \hline
    \multirow{2}{*}{Tasks} &\multicolumn{2}{c|}{Li \textit{et al.} \cite{Li_Li_Li_Zhang_Qian_2023}}   &  \multicolumn{2}{c}{PUSNet}  \\ 
    \cline{2-5} 
    &PSNR(dB)&SSIM&PSNR(dB)& SSIM\\
    \hline
    Secret embedding & - & - & 39.09 & 0.9772   \\
    Secret recovery & 28.52 & 0.8487 & 26.96 & 0.8211  \\
    Image denoising $\searrow$ & 1.24 & 0.0219 & 0.73& 0.0148  \\
    \hline
    \end{tabular}}
    \vspace{-0.3cm}

\label{tab:csm}
\end{table}

\subsection{Comparison against the SOTA}
In this section, we compare our PUSNet against the SOTA method proposed in \cite{Li_Li_Li_Zhang_Qian_2023}. Since the SOTA method is tailored for hiding a secret decoding network, we only take the decoding network of a popular DNN-based steganographic scheme (i.e., HiDDeN) as the secret DNN model for evaluation. We embed it into a benign DNN model using the SOTA method to form a stego DNN model, where the benign DNN model is with the same architecture as the secret DNN model. Table. \ref{tab:csm} gives the performance of the secret embedding and recovery tasks using the secret DNN model extracted from the stego DNN model or triggered from our purified network, where ``-" means not applicable and the secret recovery task is evaluated on the COCO dataset \cite{lin2014microsoft}. We can see that the performance of the secret recovery using the decoding network triggered from the PUSNet (i.e., the PUSNet-D) is slightly lower than that of the SOTA method. However, it brings a lower performance degradation on the image denoising task for the benign DNN model. We would also like to point out that our proposed method is able to conceal both the secret encoding and decoding networks in one single DNN model, which is much more useful than the SOTA method in real-world applications.

\section{Conclusion}
In this paper, we propose PUSNet to tackle the problem of covert communication of steganographic networks. The PUSNet is able to conceal secret encoding and decoding networks into a purified network which performs an ordinary machine learning task without being noticed. While the hidden steganographic networks could be triggered from the purified network using a specific key owned by the sender or receiver. To enable flexible switching between the purified and steganographic networks, We construct the PUSNet in a sparse weight filling manner. The switching is achieved by filling some key controlled and randomly generated weights into the sparse weight locations in the purified network. We instantiate our PUSNet in terms of a sparse image denoising network, a secret image encoding network, and a secret image decoding network. Various experiments have been conducted to demonstrate the advantage of our proposed method for covert communication of the steganographic networks.

{
    \small
    \bibliographystyle{ieeenat_fullname}
    \bibliography{main}
}


\end{document}